\documentclass[twoside]{article}

\usepackage{lipsum} 
\usepackage{graphicx}
\usepackage[sc]{mathpazo} 
\usepackage[T1]{fontenc} 
\usepackage{microtype} 
\usepackage{authblk}
\usepackage[hmarginratio=1:1,top=32mm,columnsep=20pt]{geometry} 
\usepackage[hang, small,labelfont=bf,up,textfont=it,up]{caption} 
\usepackage{booktabs} 
\usepackage{float} 
\usepackage{rotating} 
\usepackage{epstopdf}
\usepackage{lettrine} 
\usepackage{paralist} 

\usepackage{abstract} 

\usepackage{titlesec} 
\renewcommand\thesection{\Roman{section}}
\titleformat{\section}[block]{\large\scshape\centering}{\thesection.}{1em}{} 
\usepackage{hyperref}
\usepackage{fancyhdr} 
\title{\vspace{-15mm}\fontsize{24pt}{10pt}\selectfont\textbf{Comparative studies of Population Synthesis Models in the frame work of modified Str\"{o}mgren filters}} 

\author[1]{Yuvraj Harsha Sreedhar\thanks{yuvrajharsha@gmail.com}}
\author[]{Karl Rakos (late)}
\author[]{Gerhard Hensler}
\affil[]{Institute of Astronomy, University of Vienna, Vienna, Austria}

\date{}

\begin{document}

\maketitle 

\thispagestyle{fancy} 


\begin{abstract}
Evolutionary models form a vital part of stellar population research to understand their evolution, but despite their long history of development, they often misrepresent and misinterpret the properties of stellar population observed through broadband and spectroscopic measurements. With the growing numbers of these synthesis models, model comparison becomes an important analysis to choose a suitable model for upgrade. Along with the model comparison, we reinvestigate the technique of modified Str\"omgren photometry to measure reliable parameter-sensitive colours and estimate precise model ages and metallicities. The assessment of Rakos/Schulz models with GALEV and Worthey's Lick/IDS model find smaller colour variation: \textit{$\Delta$ (uz-vz)} $\leq$ 0.056, \textit{$\Delta$(bz-yz)} $\leq$ -0.05 and \textit{$\Delta$(vz-yz)} $\leq$ 0.061. The study conveys a good agreement of GALEV models with the modified Str\"omgren colours but with poor UV model predictions with the observed globular cluster data, while the spectroscopic models perform badly due to the use of older isochrone and stellar spectral libraries with inaccurate/insufficient knowledge of various stellar phases and their treatment. Overall, the assessment finds modified Str\"omgren photometry well suited to study different types stellar populations by mitigating the effects of the age-metallicity degeneracy.
\end{abstract}



\section{Introduction}

Stellar Population Synthesis models (PSMs) -- a vital tool to understand stellar systems, like globular clusters, galaxies -- provide insights to star-formation history (SFH), stellar metallicity (Z) and individual elemental abundance pattern, stellar initial mass function (IMF), total mass in stars, the dust and gas content. These properties are extracted from the spectral energy distribution (SED) of stellar systems, formed from the library of theoretical isochrones (with signature L, $T_{eff}$, initial chemical composition and mass M of individual stars) and the library of stellar spectra with the aid of stellar evolutionary theory. To this integrated spectrum corrections are applied, if needed, due to dust, nebular emission, K-correction. Eventually magnitudes/colours, chemical composition based on the line strength of the indices are measured from the resulting spectrum (Salaris et al., 2009) which relate to different scenarious of the evolution of galaxies and galaxy clusters which cannot be resolved to their individual stellar types. 

The history of stellar population modelling dates back to the attempt by Crampin \& Hoyle (1961) by correlating the observed B-V vs. $M_{V}$ diagram for various ages by assuming a constant Red Giant Branch (RGB) tip and homology of isochrones. With the first set of wide evolutionary tracks, Tinsley (1968) introduced the first synthesis models, while Spinrad \& Taylor (1971) reproduced the observed integrated light of stellar populations by combining by trial-and-error method the stellar spectra of different classes. Tinsley (1972, 1973) detected the spectrophotometric properties, approximate analytical solutions to star formation rates (SFRs), chemical compositions, IMFs to give an evolutionary perspective from visual and near-Infra Red (NIR) colours. Models by Gunn et al. (1981), which included empirical RGB, Asymptotic Giant Branch (AGB), lower-Main Sequence (MS) phases and isochrones with uncalibrated mixing length of different stellar phases, badly failed to reproduce the MS stars of the galactic Globular Clusters (GCs) and miscalculated the ages of ellipticals (by a few gigayears), this took population synthesis modelling back to the drawing board. Models developed, since then, focussed meticulously on co-evolving, chemically homogenous stars -- the so called Single Stellar Populations or SSPs -- including isochrones with calibrated mixing lengths and accounting proper energy contribution of different stellar phases. Also, in this era, the models spanned a wider range in age and metallicity to include Horizontal Branch (HB), AGB, Post-AGB phases. By 90s, isochrone synthesis technique had become popular, which led models by Bruzual \& Charlot (1993), Charlot \& Bruzual (1991) to refine colour variations and include the semi-empirical analysis of Thermally Pulsating-AGB (TP-AGB) phase. This advancement resulted in ages to extend to below 1 Gyr (for Z=$Z_{\odot}$), and paved a platform for the new generation of isochrone synthesis by Geneva (Tantalo et al. 1996) and Padova (Leitherer et al. 1999). Subsequently, other models were also developed by Worthey (1994; W94 henceforth), Worthey et al. (1994), Vazdekis et al. (1997) and Vazdekis (1999). These models set a new trend to offer a whole suit of features, for e.g. colours, spectral energy distributions, mass-to-light ratios, spectral indices, surface brightness fluctuations, metallicities, ages, IMFs, etc. These types of models were termed as the Comprehensive models (Maraston 2003).

\begin{figure}
\includegraphics[scale=0.65, angle=-90]{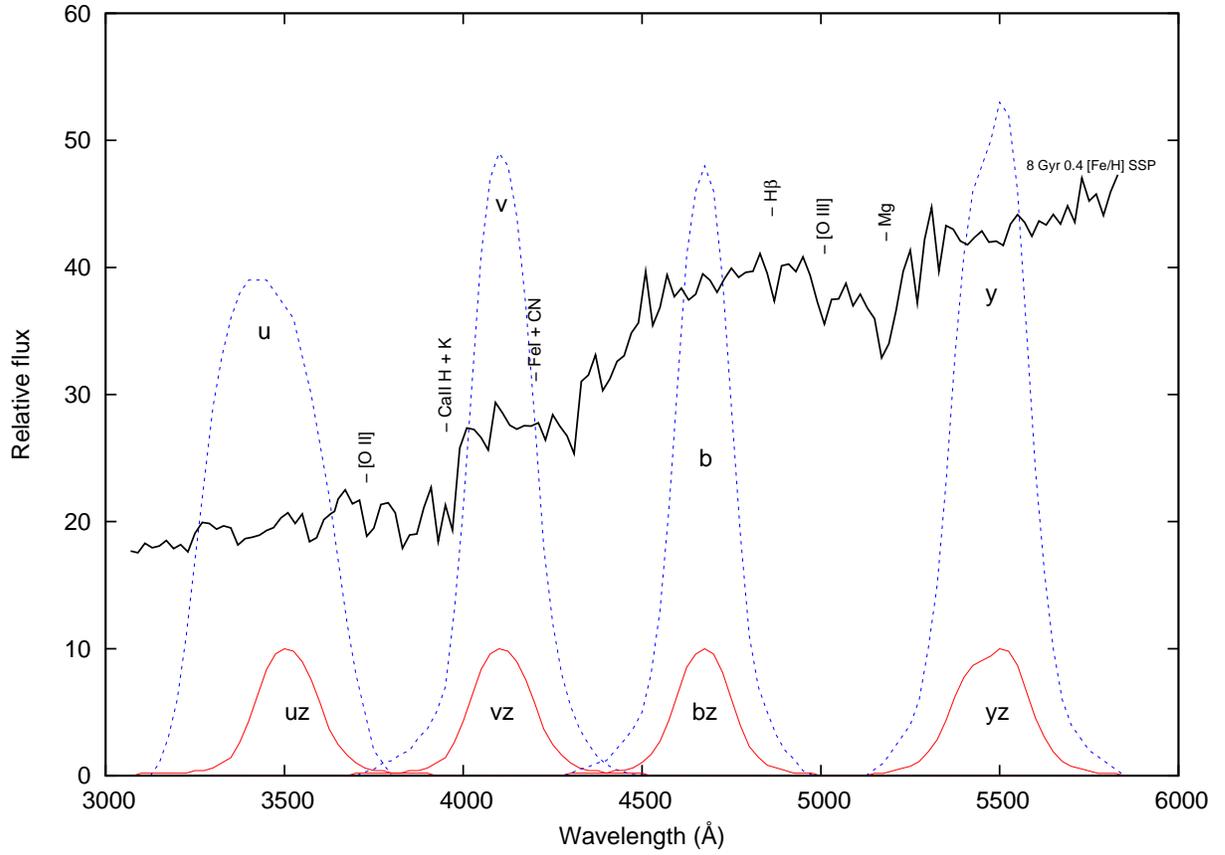}
\caption{The two sets of filters, Str\"{o}mgren and modified Str\"{o}mgren are displayed in dashed and solid lines, respectively with the 8Gyr. 0.4 [Fe/H] SSP spectra from W94 in the background. The four filters of either systems are concentrated at \textit{u (3500 \AA), v (4100 \AA), b (4675 \AA), y (5500 \AA)} to investigate the narrow regions of the spectra, essential for studying specific parameters of stellar population. Modified Str\"{o}mgren are on average narrower (20 \AA) than the conventional Str\"{o}mgren filters and are redshifted to the target cluster.}
\end{figure}

Now there are a myriad number of stellar population models in the scientific community, 
amongst which, the latest classification (Chen et al. 2010) includes models with the library of empirical spectra of stars and star clusters, termed as the empirical  population synthesis models (Faber 1972; Bica 1988; Boisson et al. 2000; Cid Fernandes et al. 2001), and, the other is the conventional, evolutionary population synthesis models (EPS; Tinsley 1978; Bruzual 1983; Worthey 1994; Leitherer \& Heckman 1995; Maraston 1998; Vazdekis \& Arimoto 1999; Bruzual \& Charlot 2003; Maraston 2005; Cid Fernandes et al. 2005) which uses the knowledge of stellar evolution to model the spectrophotometric properties of stellar populations. In the empirical models, the observed spectrum of a galaxy is reproduced by a combination of spectra of individual stars and stellar populations with different ages and metallicities from a library. Whereas, in the EPS, the spectrum of an individual stellar population or galaxy is formed by the combination of several stellar population spectra by adjusting parameters, like, the stellar evolution tracks, the stellar spectral library, the IMF, the SFH and the grids of ages and metallicities. 

These PSMs have greatly benefited our understanding of stellar structures as a function 
of time, chemical composition, IMF. Besides, their paramount importance of reproducing the integrated light of resolved SSPs, they are also important in interpreting unresolved stellar populations with unknown SFH, if the model hosts a bigger archive of SSP spectra covering a large range of ages and metallicities, (like galaxies, starbursts, galaxy groups and clusters) with the linear combination of input spectra (Schulz et al. 02; S02 henceforth). PSMs are also helpful in understanding the cosmological structure formation scenario, if individual galaxies were built from the sub-galactic fragments to a state resembling the Milky Way in several aspects (Contrado et al. 1999).

However, despite decades of history in developing and improving of these PSMs, several critical caveats are still observed that constrain correct measurement of stellar population 
properties, like, 1. the behaviour and the contribution of several stellar evolutionary  phases, like, the TP-AGB, blue-stragglers, convective core overshoot, helium abundance 
(specially at higher metallicities), etc., are not well known, their effects create enormous ($\sim$ 1-2 mag) model divergence from the observed -- a serious issue to be corrected in isochrones. 2. almost all spectral libraries archive spectra of low resolution which result in an uncertain estimation of chemical composition and age. In order for these models to measure the strength of numerous weak absorption lines and strong lines over a wide range of ages, high resolution spectral libraries are urgently needed. 3. The study of galaxy formation and SFH, along with its age, would clearly be based on observations of nearby galaxies (Worthey 1994), but the Z effect masks the age effect. This results in erroneous age and/or Z measurements, the so called age-metallicity degeneracy (Worthey 1999). A model which can break the degeneracy to give precise ages and metallicities is required. 4. The limited range of the parameters and their combinations, especially of age and Z, introduces problems in linear interpolation and extrapolation to the higher and lower orders. This range needs to extended. 5. Salaris et al. (2009) point out that the lack of a single theoretical (or empirical) spectral library, covering all relevant parameter space. 

In an attempt to constrain some of these errors, flaws and model divergence, Rakos \& Schombert (2005a; RS05a henceforth) developed a semi-empirical population synthesis model by incorporating Schulz et al. (2002; S02) with the application of the Principle Component Analysis (PCA; Steindling et al. 2001; S01) to suit their novel modified Str\"omgren filters (Fiala et al. 1986, Rakos et al. 1988, 1990; discussed in \S 2.3). Prior to adopting a new set of improved models (to further resolve model errors), as an objective of this paper, we perform a model comparison study of RS05 with, one which is fairly new (2009) and, the other, that is carefully studied using the Lick indices. This assessment aims to check for reliability and reproducibility of the modified Str\"omgren colours, to compare with the different models, and re-investigate the filter transmission curves of the modified Str\"omgren filters by the agreement of convolved colours with those of model SSPs. This would verify the technique's ability to study stellar populations properties of single and composite nature. 

Although there are several other, which are much more improved and comprehensive EPS and empirical models, but in order to keep our current study straightforward and vital, we choose to probe with the established and thoroughly analysed models. Nevertheless, new models, like the Flexible Stellar Population Synthesis (FSPS; Conroy et al. 2009, 2010) and STARLIGHT (Cid Fernandes 2007, 2013), are currently being assessed and would be addressed in our subsequent paper. But for our current study, as a fairly new model, we opt for GALEV (Kotulla et al. 2009; K09) -- successor of S02 models -- which include new features of latest evolutionary tracks, better treatment of TP-AGB and convective overshooting phases while offering a more comprehensive output (discussed in \S2.1).  And, from the Lick/IDS study perspective, we choose W94 (elaborated in \S2.2) models which are based on reliable input stellar evolutionary isochrones from VandenBerg and collaborators and the Revised Yale isochrones (Green et al. 1987) with proper accounting of errors in the estimation of age, metallicity, IMF (W94). For this comparison, we arrange this paper with the Comparison Components describing Schulz and GALEV models in \S 2.1, the Lick/IDS systems in \S 2.2, Rakos's modified Str\"{o}mgren filter systems \S 2.3, the comparison method of models in \S 3. Finally in \S 4, we reiterate with conclusions.

\section{Comparison Components}

\subsection{Schulz/GALEV Population Synthesis Models}

The Gottingen group of S02 devised this SSP based PSMs with the metallicity range of $0.02 \leq Z/Z_{\odot} \leq -2.5$ and age range of $4 \times 10^6$ yr $\leq t \leq 16$ Gyr. They use the Padova group isochrone library with TP-AGB stars in the range of $2M_{\odot} \leq$ M $\leq 7M_{\odot}$ and stellar atmosphere spectral library from Lejeune et al. (1997, 1998). Lejeune et al.'s single burst star-formation model libraries are in the wavelength range of $90$ \AA\ to $160$ $\mu$m, which they claim to be complete in terms of stellar effective temperature from $T_{eff}$ =2800 K to $T_{eff}$ =47, 500 K, and surface gravities from $-1.0 \leq log$ g $\leq 5.5$. 

However, poor resolution of S02's spectra are of the order $20$\AA\ in the wavelength range $3000$\AA\ to $10000$\AA\ and beyond that, even worse, to $50$\AA. This poor resolution hinders an in-depth analysis of individual line indices. Spectral comparison between various models shows very poor agreement for wavelengths below $2000$\AA, where the major contributors are hot stars and cool white dwarfs. Another drawback in this PSM comes from the use of broadband colours in measuring age and metallicity, which are highly contaminated with age-metallicity degeneracy (Worthey 1999). This would result in colour discrepancy between the model spectra and the observed values. Most PSMs, including S02, have a limited range of Z values based on the Milky Way or M31 GCs. 

GALEV (K09) offers users many features, one of which is its chemically consistent models, such that a stellar spectral evolution could be studied with the evolution of chemical composition of Interstellar Medium (ISM). Yet another GALEV's ingenious feature is its very informative output on spectra, emission and absorption line indices, photometric magnitudes/colours for a range of filter systems, chemical abundances, gaseous and stellar masses, SFR in their time evolution of normal, starburst galaxies and galaxies with truncated SFR.  However, they still use the same spectral and isochrone libraries as S02 models and, hence, remains the poor resolution of $20$\AA\ for UV-optical and $50-100$\AA\ in the NIR (Near-IR) wavelength.

Even after such careful up-gradation from S02 to GALEV, there are still discrepancies in the observed and the modeled spectra. From Figure 8 in K09, that displays the comparisons of model spectra with the local templates for different (Hubble) galaxy types from the Kennicutt (1992) catalogue, one could notice significant differences in each galaxy type spectra amongst differently coloured templates. Agreeably, these differences are indeed small, with an average maximum of the relative difference of 0.3 (refer Figure 8 in K09) for most galaxy types upto $\lambda$=4500\AA, which comprise two of our special Str\"{o}mgren filters, namely \textit{uz} (3500 \AA), \textit{vz} (4100 \AA). Similar order differences are also observed in our colour estimations of modified Str\"{o}mgren photometry (discussed in \S 3.3).

\begin{table}[htbp]
\caption{The comparison presents the colours of different SSP models of different ages and metallicities of the modified Str\"{o}mgren colours between Rakos (RS05) with GALEV (K09) and Spectroscopic Lick/IDS (W94) studies.}
\centering
\begin{tabular}{ccccccccccc}
\textbf{} & \textbf{} & \textbf{RS05} & \textbf{RS05} & \textbf{RS05} & \textbf{GALEV} & \textbf{GALEV} & \textbf{GALEV} & \textbf{W94} & \textbf{W94} & \textbf{W95} \\ 
\textbf{Z} & \textbf{Age} & \textbf{(uz-vz)} & \textbf{(bz-yz)} & \textbf{(vz-yz)} & \textbf{(uz-vz)} & \textbf{(bz-yz)} & \textbf{(vz-yz)} & \textbf{(uz-vz)} & \textbf{(bz-yz)} & \textbf{(vz-yz)} \\ 
dex & Gyr &  &  &  &  &  &  &  &  &  \\ 
 &  &  &  &  &  &  &  &  &  &  \\ 
-1.7 & 3 & 0.73 & 0.12 & 0.03 & 0.71 & 0.13 & 0.03 &  &  &  \\ 
-1.7 & 4 & 0.69 & 0.14 & 0.07 & 0.66 & 0.15 & 0.08 &  &  &  \\ 
-1.7 & 6 & 0.65 & 0.17 & 0.12 & 0.61 & 0.18 & 0.14 &  &  &  \\ 
-1.7 & 8 & 0.62 & 0.18 & 0.14 & 0.58 & 0.19 & 0.15 & 0.596 & 0.144 & -0.090 \\ 
-1.7 & 10 & 0.6 & 0.19 & 0.16 & 0.57 & 0.2 & 0.17 & 0.592 & 0.162 & -0.052 \\ 
-1.7 & 12 & 0.61 & 0.19 & 0.17 & 0.56 & 0.21 & 0.18 & 0.588 & 0.177 & -0.021 \\ 
-1.7 & 14 & 0.62 & 0.2 & 0.18 & 0.58 & 0.21 & 0.18 & 0.591 & 0.189 & 0.006 \\ 
-0.7 & 3 & 0.72 & 0.21 & 0.31 & 0.71 & 0.21 & 0.3 &  &  &  \\ 
-0.7 & 4 & 0.72 & 0.22 & 0.32 & 0.69 & 0.22 & 0.32 &  &  &  \\ 
-0.7 & 6 & 0.71 & 0.23 & 0.36 & 0.68 & 0.24 & 0.37 &  &  &  \\ 
-0.7 & 8 & 0.73 & 0.26 & 0.43 & 0.7 & 0.26 & 0.43 & 0.659 & -0.040 & -0.074 \\ 
-0.7 & 10 & 0.73 & 0.26 & 0.44 & 0.71 & 0.27 & 0.44 & 0.669 & 0.181 & 0.163 \\ 
-0.7 & 12 & 0.76 & 0.28 & 0.49 & 0.73 & 0.28 & 0.49 & 0.677 & 0.362 & 0.356 \\ 
-0.7 & 14 & 0.76 & 0.27 & 0.48 & 0.73 & 0.28 & 0.49 & 0.693 & 0.501 & 0.511 \\ 
-0.4 & 3 & 0.74 & 0.23 & 0.35 & 0.71 & 0.23 & 0.37 &  &  &  \\ 
-0.4 & 4 & 0.72 & 0.23 & 0.38 & 0.69 & 0.23 & 0.37 &  &  &  \\ 
-0.4 & 6 & 0.73 & 0.26 & 0.44 & 0.72 & 0.26 & 0.44 &  &  &  \\ 
-0.4 & 8 & 0.77 & 0.28 & 0.51 & 0.75 & 0.28 & 0.52 & 0.686 & 0.135 & 0.208 \\ 
-0.4 & 10 & 0.79 & 0.29 & 0.54 & 0.76 & 0.29 & 0.54 & 0.701 & 0.357 & 0.450 \\ 
-0.4 & 12 & 0.8 & 0.29 & 0.54 & 0.78 & 0.29 & 0.55 & 0.714 & 0.539 & 0.647 \\ 
-0.4 & 14 & 0.83 & 0.3 & 0.59 & 0.81 & 0.31 & 0.59 & 0.731 & 0.688 & 0.814 \\ 
0 & 3 & 0.75 & 0.26 & 0.46 & 0.75 & 0.26 & 0.46 & 0.706 & 0.302 & 0.422 \\ 
0 & 4 & 0.81 & 0.29 & 0.58 & 0.8 & 0.29 & 0.57 & 0.707 & 0.304 & 0.438 \\ 
0 & 6 & 0.78 & 0.29 & 0.57 & 0.81 & 0.29 & 0.57 & 0.723 & 0.318 & 0.489 \\ 
0 & 8 & 0.87 & 0.31 & 0.65 & 0.85 & 0.31 & 0.64 & 0.746 & 0.337 & 0.550 \\ 
0 & 10 & 0.89 & 0.32 & 0.67 & 0.88 & 0.33 & 0.68 & 0.765 & 0.347 & 0.587 \\ 
0 & 12 & 0.95 & 0.34 & 0.74 & 0.94 & 0.35 & 0.75 & 0.780 & 0.356 & 0.617 \\ 
0 & 14 & 1 & 0.35 & 0.78 & 0.97 & 0.35 & 0.78 & 0.797 & 0.363 & 0.644 \\ 
0.4 & 3 & 0.87 & 0.29 & 0.6 & 0.86 & 0.29 & 0.6 & 0.896 & 0.335 & 0.616 \\ 
0.4 & 4 & 0.95 & 0.33 & 0.74 & 0.93 & 0.33 & 0.73 & 0.946 & 0.361 & 0.696 \\ 
0.4 & 6 & 1.01 & 0.33 & 0.74 & 1 & 0.34 & 0.79 & 1.006 & 0.388 & 0.781 \\ 
0.4 & 8 & 1.06 & 0.36 & 0.84 & 1.06 & 0.36 & 0.85 & 1.039 & 0.398 & 0.817 \\ 
0.4 & 10 & 1.11 & 0.38 & 0.88 & 1.1 & 0.38 & 0.9 & 1.090 & 0.409 & 0.864 \\ 
0.4 & 12 & 1.13 & 0.38 & 0.89 & 1.14 & 0.38 & 0.92 & 1.131 & 0.419 & 0.902 \\ 
0.4 & 14 & 1.21 & 0.39 & 0.96 & 1.19 & 0.4 & 0.96 & 1.161 & 0.424 & 0.923 \\ 
 &  &  &  &  &  &  &  &  &  &  \\ 
 &  &  &  &  &  &  &  &  &  &  \\ 
\end{tabular}
\label{}
\end{table}


\subsection{Spectroscopic Lick Indices}

An extensive work on these PSMs has been done by W94, which includes comparison of modeled line indices and observed spectroscopic methods. W94 points out that these models assume exactly one age and one Z for an entire stellar population, hence, one-to-one comparison of galaxies is not possible. In reality, however, galaxies are composite in Z (at least) and age. Furthermore, models also assume that the HB stars remain in the red clump of the giant branch, but that is not the case for metal poor stars. Observations have confirmed that galaxies with metal poor stars indeed show extended horizontal branches. However, models that could compare and successfully match the observed phenomenon in CSPs are eagerly awaited. 

Age-metallicity degeneracy is a long standing problem in galaxy evolutionary studies and W94 suggests that one possible way to separate age effects from metallicity effects, is by using respectively sensitive colours and/or line indices. Any variation in a particular line index by an amount \textit{$\Delta$I} (refer Table 6 in W94) could be explained by either due to change in age or change in metallicity. W94 also conveys that the most commonly used Z  and age sensitive line indices are often contaminated by indices of other elements.


\subsection{Modified Str\"{o}mgren Photometry}

In principle, the modified Str\"omgren filters, denoted as uz, vz, bz, yz (where z represents their redshifted nature), are not much different from the normal Str\"omgren (u, v, b, y) filters, as the effective wavelengths of the two filter systems are almost same. However, the modifications in the modified Str\"omgren filters come from the following aspects. (1) The uz filter is slightly (30\AA) shifted to the red to focus more on the red galaxies. (2) The calibration of these filters is performed using spectrophotometric standard stars which are shifted to the redshift of the target cluster (Rakos et al. 1988). This makes the zero point of the magnitudes differ along with the cluster's redshift. (3) The flux measurement for the galaxies and the standard stars is in wavelength units (per \AA), as is usual for most galaxy studies instead of frequency (per Hz) units -- as in the stellar studies (Sreedhar et al. 2012).

These modified Str\"{o}mgren filters were initially designed by Fiala et al. (1986) and later used by Rakos et al. (1995) to study galaxies in clusters. Figure 1 shows the similarity of Str\"{o}mgren and modified Str\"{o}mgren filter transmission curves. The modified Str\"omgren filter system covers three regions in the near-UV and one in the optical blue portion of the electromagnetic spectrum. The \textit{bz} \textit{($\lambda_{eff}$=}4675\AA) and \textit{yz} \textit{($\lambda_{eff}$=}5500\AA) focuses on the continuum part of the spectrum, both of which combine to form the temperature-colour index \textit{(bz-yz)}. Filter \textit{vz} \textit{($\lambda_{eff}$=}4100 \AA) is strongly influenced by the metal absorption lines (i.e Fe, CN) from  old stellar populations, whereas \textit{uz} \textit{($\lambda_{eff}$=}3500\AA) is short ward of the Balmer jump.

These filters are narrow enough to ensure the spectral purity and the passbands which are well placed to study the Balmer discontinuity and the spectral continuum covered by our \textit{bz} and \textit{yz} filters. The idea is to look for the change in colour of the RGB, that acts as a metallicity indicator, and similarly change in colour, produced by the shifting the turnoff points, reflects the age (Tinsley et al. 1980) in galaxies.

\newpage
\begin{figure}
\centering{
\includegraphics[scale=0.7, angle =-90]{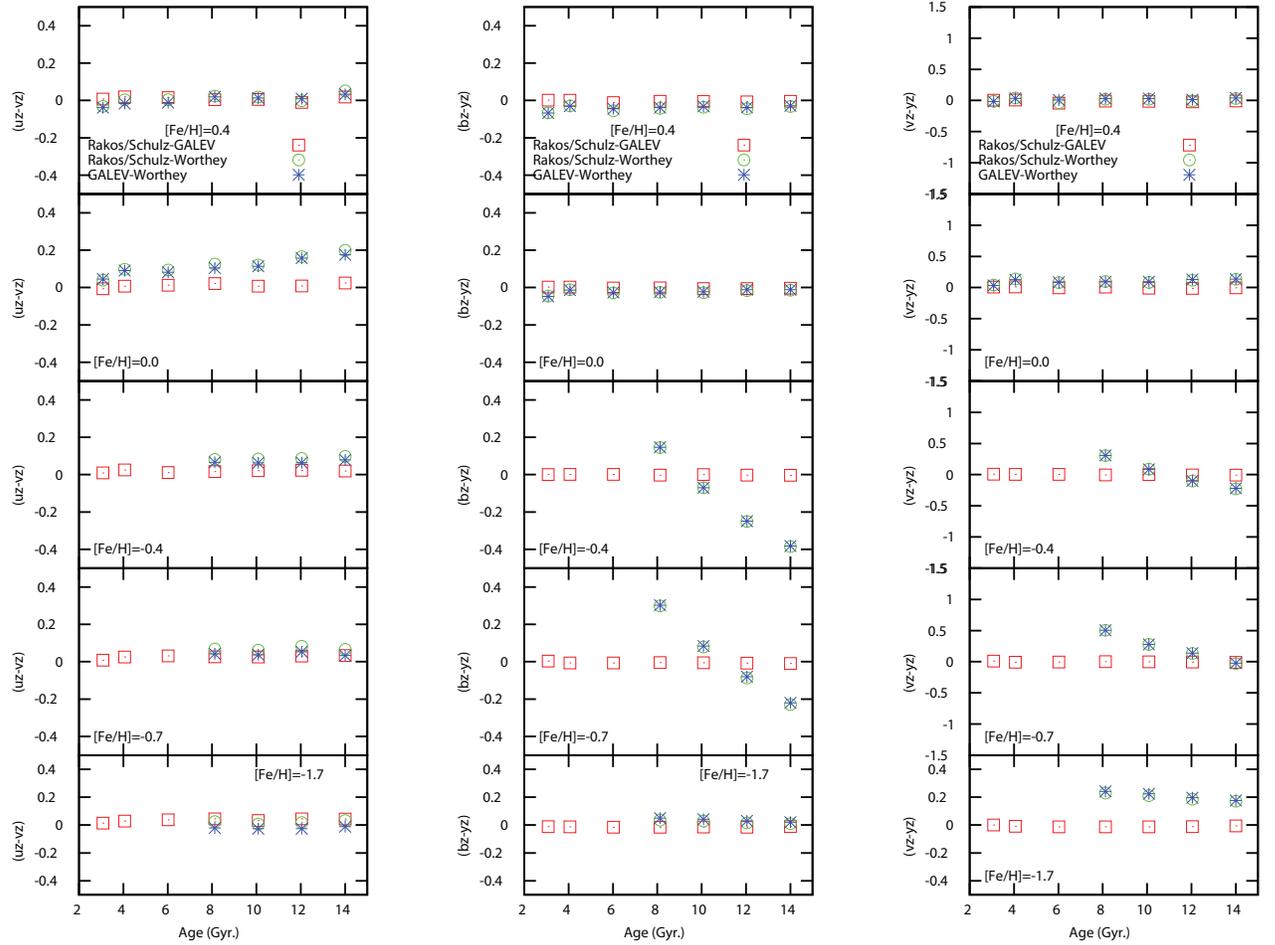}
\caption{Above plots show the difference in modified Str\"{o}mgren colours between the  RS05 model with GALEV and W94 for different Zs and ages. Difference between Rakos/Schulz-GALEV are shown in squares, Rakos/Schulz-Worthey in circles, GALEV-Worthey in asterisks. Comparison shows close agreement of Rakos/Schulz with GALEV, however discrepancy between GALEV-Worthey and Rakos/Schulz-Worthey are observed to be large.}}
\end{figure}

Since broadband colours are not well suited for separating age effects from metallicity effects, Rakos et al. (1995, 2001, 2005, 2007) attempted to solve the age-metallicity degeneracy problem uniquely by using narrow-band colours with the PCA technique. The PC analysis is a three-dimensional multicolour space defined by these modified Str\"{o}mgren colours and is formed by three PC equations (refer S01), where different Hubble type galaxies take selected places in that cluster-box. 

Acceptable linearity for ages greater than 3 Gyr over a full range of metallicities, needed for application of these PC analysis, were found with S02 models (RS05). Interpolation between the model grids were performed by the well studied GCs (shown in Figure 3) using modified Str\"omgren photometry, hence referred as the semi-empirical models. The original S02 models offer a carefully calibrated Str\"{o}mgren magnitudes (Gray et al. 1998) for a set of metallicity values for a range of ages, that are transformed to modified Str\"{o}mgren filter system using the following equations from Rakos et al. (1996):

\begin{equation}
      \textit{(bz-yz)} = -0.268 + 0.973 \textit{(b-y)}
\end{equation} 
\begin{equation}
     \textit{mz} = 1.092 m_{1} - 0.017 \textit{(b-y)} 
\end{equation} 
\begin{equation}
     \textit{cz} = 0.234 + 1.034 c_{1} - 0.152 \textit{(b-y)} 
\end{equation}

where, $m_{1}$ and $c_{1}$ are the old Str\"{o}mgren metal-line and surface gravity indices given by 

\begin{equation}
m_{1} = (v-b) - (b-y)
\end{equation} 
\begin{equation}
c_{1} = (u-v) - (v-b)
\end{equation}

These modified Str\"{o}mgren colours are tabulated in RS05 and here in columns 3 - 5 of Table 1 for comparison. The ages and metallicities estimated using these semi-empirical models are found to have errors of 0.2 dex in Z and 0.5 Gyr in age. 

Acknowledging the inaccurate and incorrect model estimates, Rakos and co-workers took careful measures in order to prevent and/or curb these model divergence benefiting these modified Str\"omgren colour indices which show sensitivity towards metallicity, dust, 4000 \AA\ break and age of the underlying populations, which are discussed in different articles. In Rakos et al. (1990), they draw a direct, empirical relation to [Fe/H] (between -2.1 $<$ [Fe/H] $<$ 0) using (vz-yz), (bz-yz) for SSPs and CSPs. In Rakos et al. (2001), this relation was updated for a much tighter one using the vz-yz index with observations of the 41 Milky Way GCs and Fornax dwarf ellipticals (using spectroscopic measurements by Held \& Mould 1994). In Rakos \& Schombert (2004), this updated relation is compared and found to fit well with Schulz et al. (2002) model metallicities in a colour-colour diagram with M87 globulars. Whereas, the uz-vz index correlation to the amplitude of spectral [D(4000\AA)] -- a parameter-sensitive to the metal content -- is shown in Rakos et al. (2001). The residual (bz-yz) shows a good fit with Milky Way globulars and derived isochrone ages from Salaris \& Weiss (1998) which is illustrated in Rakos et al. (2004). The PCA photometric ages are also compared with Schulz et al. (2002) model ages in Rakos et al. (2005a). In addition, Rakos et al. (2001) outline a relation of vz-yz index with $Mg_2$ abundance to find that with increasing vz-yz colour, there is a decrease in [Mg/Fe] ratio -- an indirect measure of star formation history. S01 point out that since the reddening vector forms at a large enough angle from the effects of age and metallicity, this feature could alleviate the effects of age-dust-metallicity degeneracy (Worthey 1994) by employing PCA technique using rest-frame narrowband colours (Sreedhar [submitted]).


\section{Comparison of SSP colours between models and spectroscopy with respect to modified Str\"omgren filter system}

The following comparisons are between the semi-empirical and theoretical models using SSP spectra of different ages and metallicities. Previously estimated colours by RS05 using S02 models (are shown as Rakos/Schulz or RS05), as discussed earlier are compared with the S02's latest upgrade, GALEV (K09)\footnote{S02 and the latest GALEV models are obtained from the website: www.galev.org (Kotulla, private communication)}. Also, in the comparison are the colours from W94's\footnote{Spectra can be obtained from http://astro.wsu.edu/worthey/dial/dial\_a\_model.html} spectra which are convolved using modified Str\"{o}mgren filter transmission curves. 

GALEV colours in the modified Str\"omgren filter system are estimated by converting from the Str\"omgren magnitudes using eqns.(1)-(3), and applying the zero-point offsets for the Vega system from Gray et al. (1998). The GALEV models offer SSPs for 5 different metallicity values (-1.7, -0.7, -0.3, 0, 0.3 dex) and 6 different ages (3.12, 4.06, 6.02, 10.08, 12.04, 14 Gyr). These GALEV colours are shown in columns 6 - 8 of Table 1. One important aspect to note here is the small change in metallicity values of $\pm0.3$ dex and $\pm0.4$ dex brought into GALEV from that of S02 models, respectively; this may affect only very negligible colour difference.

Spectra from W94 of similar ages and metallicities as Rakos/Schulz and GALEV are convolved through the modified Str\"{o}mgren filters transmission curves to obtain their colours, using the eqns.(1)-(3), in the modified Str\"{o}mgren filter system; these are shown in column 9 - 11 of Table 1. Blank values in these columns indicate the non-availability of some young, low metallicity SSP spectra -- this could be, perhaps, due to the G dwarf problem (van den Bergh 1962, Schmidt 1963). 

\begin{figure}
\centering
\includegraphics[scale=1.3]{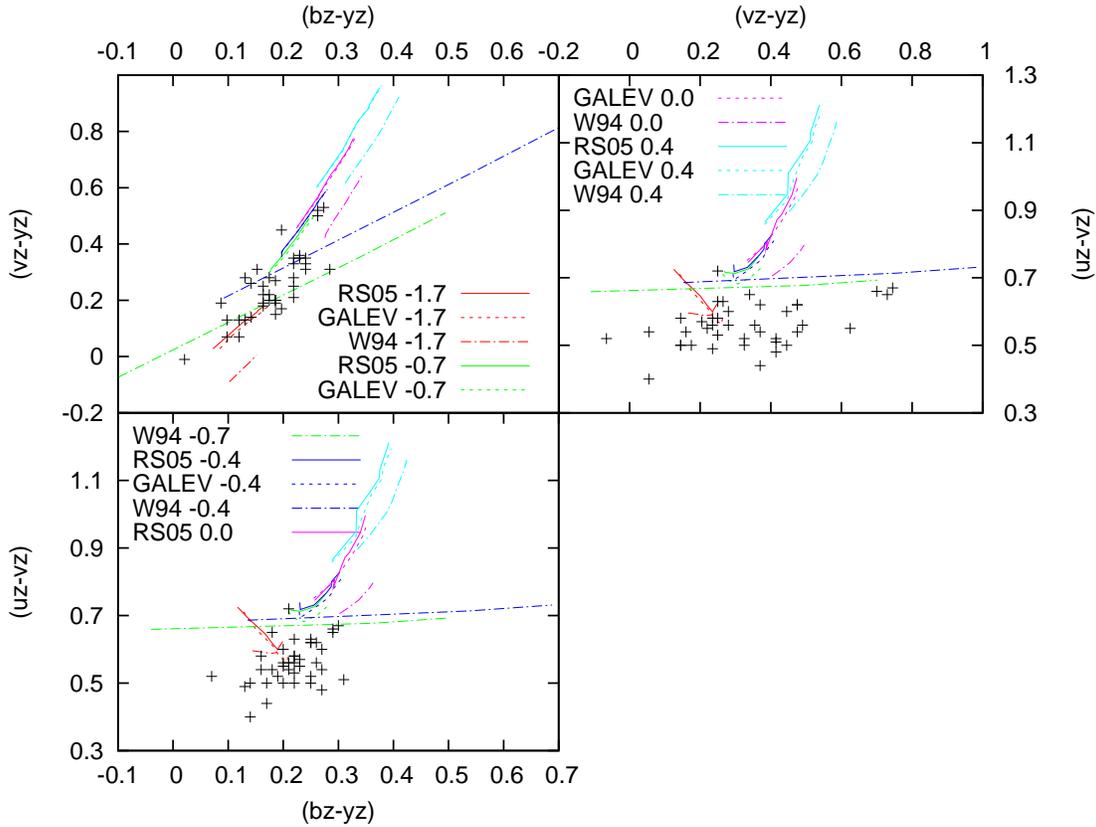}
\caption{Model colour tracks with observed Globular Clusters. Modified Str\"omgren colour tracks of different model metallicities and ages fro RS05, GALEV, Worthey are displayed and labelled in three different plots. Observed Globular Clusters by RS05 are also illustrated (these GCs are tabulated in Table 2 in RS05). Clear similarities between the RS05 and GALEV can be seen, while the spectroscopic models by W94 are observed to run in different direction. Also, the GC data tend to show a very good agreement with vz-yz and bz-yz model colours of RS05 and GALEV, while uz-vz model colours are much bluer than the observed.}
\end{figure}

The table shows close similarities in colours of different ages and metallicities observed between the RS05 and GALEV to testify the reliability of these equations (1)-(3) which transform Str\"{o}mgren to modified Str\"{o}mgren colours. While the colours of the W94 spectra show some offset from the RS05 and GALEV models. These offset values between individual model colours, for varying Z and age, are illustrated in Figure 2. Although, most metallicity models, displayed in the figure, show points close to the zero mark to convey least offset values. The minimum values in these colour differences are found between the RS05 and GALEV models, whereas the maximum are observed between RS05-Worthey and GALEV-Worthey comparisons. Overall comparison between the three models shows the colour variation as: 0.017$\leq$ \textit{$\Delta$ (uz-vz)} $\leq$ 0.056, -0.004 $\leq$ \textit{$\Delta$(bz-yz)} $\leq$ -0.05 and -0.004 $\leq$ \textit{$\Delta$(vz-yz)} $\leq$ 0.061.  

Figure 3 depicts the model colour tracks, in the modified Str\"omgren photometric system, of the three models, while the plus symbols show the observed GC colours by RS05, and the ages and metallicities presented by Harris (1996) and Salaris \& Weiss (2002); these GC data are shown in Table 2 of RS05. The RS05 and the GALEV model tracks show clear similarities, while the spectroscopic models by W94 strangely present a very different colour behaviour in all three colour plots. The observed colours of GCs are found to agree well with RS05 and GALEV models in the vz-yz and bz-yz colour plot, while being way off in the uz-vz index. Besides the reason of poor (20\AA) spectral resolution of S02 and GALEV models, as pointed out earlier, this disagreement with (uz-vz) could be due to poor knowledge and treatment of HB and Blue Stragglers stellar phases at UV wavelengths in most models, as also illustrated by Carter et al. (2009) to find the divergence in the GALEV model predictions by 0.05-0.1 mag using broadband filters (Sreedhar [submitted]). Although this sort of divergence is found to be a common problem amongst most models, and their correction is an urgent requirement. 

Overall deviations in colours of Rakos/Schulz, GALEV with Worthey could possibly be explained by the following reasons: 1. Colours could deviate by the incorrectly measured metallicities in spectra, due to the contamination of different line indices. Besides that, several other model caveats, as discussed in the introduction, are also known to affect the model colours. 2. Differences in colours could also arise due to the limited understanding of theoretical stellar atmospheres, chemical enrichment processes, which could result in the change in the spatial pattern of the spectra and, hence, the erroneous colours. 3. The redder \textit{(vz-yz)} spectral colours could be explained due to exclusion of low metallicity stars. 4. The geometrical behaviour of data sampling is also different such that slit or fiber spectroscopy that measures the surface brightness of central core regions of a galaxy, instead of integrated light of the whole galaxy as in photometry (Schombert \& Rakos 2009). This would in turn result in wrong metallicity measurements. 5. Minor differences due to the convolution of colours by the filter transmission curves and small errors due the interpolation between the model grids are also expected to create model divergence. 

From the above re-investigation, we conclude: 1. By the observed similarities in colours (specially in vz-yz and bz-yz) the interpolation/modifications performed by RS05 (using the S02 models with the PCA technique) -- to estimate precise ages and metallicities and to reduce the effect of age-metallicity degeneracy -- are in good agreement with GALEV and are found to be valid. Therefore, similar interpolation between the GALEV model grids can be utilised in understanding stellar populations. However, like most models, even GALEV present poor UV predictions for observed stellar populations, which needs to improved vastly. 2. The modification of the modified Str\"omgren filter system is well designed to accurately study the evolutionary properties of stellar populations of single and composite nature.


\section{Conclusions}

Considering the importance of PSMs in todays scientific community and their vast numbers that are available, it becomes very difficult to choose an appropriate PSM to incorporate in stellar population studies and to suit Rakos's unique (rest-frame, narrowband feature) modified Str\"{o}mgren photometric technique, that in many ways, are superior to other observational techniques. Therefore, before upgrading the models, we investigate a model comparison with a new (GALEV) and a Lick/IDS study related model (Worthey). This would relate to the technique's reliability of measuring colours and estimating precise model ages and metallicities to study stellar populations. From this study, we infer the following conclusions:
\begin{enumerate}
\item Str\"{o}mgren colours which are converted to modified Str\"{o}mgren colours using transformation equations (1-3) are found to be valid. These equations produce similar colours for different SSP model ages and metallicities.  

\item A  good agreement found between the two (RS05 and GALEV) models conveys GALEV to be suitable model upgrade replacement. This should also extend the age and metallicity precision to 0.5 Gyr and 0.2 dex, respectively, as shown in RS05 with S02 models. 

\item The modified Str\"{o}mgren colours obtained by the filter transmission convolution of W94's spectra present offsets by a few hundredths of a magnitude. These differences could be explained due to the differential geometrical observing method of spectroscopy and the redder \textit{(vz-yz)} colours, due to the exclusion of low metal stars. Also, their poor performance could be because of the use of older isochrone and stellar spectral libraries with inaccurate/insufficient knowledge of various stellar phases and their treatment. 

\item Comparison of the model colours with the observed GC data shows that vz-yz and bz-yz indices of RS05 and GALEV models are able to precisely estimate the stellar properties; the model uz-vz colours are found to be bluer than expected -- an affect due to  poor knowledge and model treatment for the HB and blue straggler stellar phases -- a rectification that is urgently required.

\item In conclusion, we find the modified Str\"omgren filter transmission curves to be well placed in the SED to measure parameter-sensitive colours, which are found to agree well with GALEV model colours to measure precise ages and metallicities. This reinvestigation finds the modified Str\"omgren photometry as a suitable technique to study stellar population properties, of single and composite nature. Latest and improved model comparisons would be assessed in the subsequent papers.

\end{enumerate}
\section{Acknowledgments}
We are grateful and extremely thankful to Andrew Paul Odell for his extensive guidance and support. We thank personally Ralf Kotulla for hosting at the University of Hertfordshire, Hatfield, UK and the entire GALEV team for helping us to understand, implement their latest models to our observational technique. We also thank Guy Worthey for sharing his spectral database. The following work was gratefully supported by the University of Vienna within the Initiative College "Cosmic Matter Circuit" (IK 538001).




\end{document}